# STATISTICS OF A FREE SINGLE QUANTUM PARTICLE AT A FINITE TEMPERATURE

JIAN-PING PENG

Department of Physics, Shanghai Jiao Tong University, Shanghai 200240, China

Abstract

We present a model to study the statistics of a single structureless quantum particle freely moving in a space at a finite temperature. It is shown that the quantum particle feels the temperature and can exchange energy with its environment in the form of heat transfer. The underlying mechanism is diffraction at the edge of the wave front of its matter wave. Expressions of energy and entropy of the particle are obtained for the irreversible process.

Quantum mechanics is the theoretical framework that describes phenomena on the microscopic level and is exact at zero temperature. The fundamental statistical character in quantum mechanics, due to the Heisenberg uncertainty relation, is unrelated to temperature. On the other hand, temperature is generally believed to have no microscopic meaning and can only be conceived at the macroscopic level. For instance, one can define the energy of a single quantum particle, but one can not ascribe a temperature to it. However, it is physically meaningful to place a single quantum particle in a box or let it move in a space where temperature is well-defined. This raises the well-known question: How a single quantum particle feels the temperature and what is the consequence? The question is particular important and interesting, since experimental techniques in recent years have improved to such an extent that direct measurement of electron dynamics is possible.[1,2,3] It should also closely related to the question on the applicability of the thermodynamics to small systems on the nanometer scale.[4]

We present here a model to study the behavior of a structureless quantum particle moving freely in a space at a nonzero temperature. As far as time evolution is concerned, we only know empirically that the process is irreversible and the particle is described by quantum mechanics originally and by statistical mechanics eventually. However, it remains so-far unanswered what is going on exactly during the intermediate process and to what extent the particle response to the temperature of its environment. Our model shows that a free quantum particle feels the temperature due to diffraction at the edge of the wave-front of its matter wave. Both quantum and statistical mechanics become imperative to describe the intermediate process of a microscopic particle at a finite temperature.

The particle is assumed to be of mass $m$ and have no internal structure. The space here at constant temperature $T$ acts only as a heat reservoir when it is in thermal contact with a thermodynamic system. We do not go into details about the space or its interaction with the particle, although the space may be filled with electromagnetic radiation just as the cosmic background radiation in the universe.

Let $E_0$ be the kinetic energy of the particle initially at the origin, the de Broglie wavelength is

$$\lambda = h/\sqrt{2mE_0} \qquad (1)$$

where $h$ is the Plank's constant. The matter wave here for the particle is assumed to a pulse or a wave-packet sharply peaked at wavelength $\lambda$. The wave front propagates along the x-axis at the phase velocity $V=\lambda E_0/h$. For simplicity, we start from a circular wave-front of radius $a_0$, which is assumed to be finite but large compared with the wavelength. According to Huygens-Fresnel principle, every point of a wave-front may be considered as a center of a secondary disturbance which gives rise to forward-going semi-spherical wavelets, and the wave-front at any later instant is the result of mutual interference of the secondary wavelets.[5] The conclusion follows directly that the diffraction angle in our model is zero, i.e., the shape and linear dimension of the forward-going wave-front in Fig.1 remains unchanged as the wave propagates. In most plane-wave related problems treated so far, $a_0$ is set to be infinity so that any effect arising from the edge of the wave-front is negligible. The assumption of a finite $a_0$ for a plane-wave here calls for a closer observation of diffraction at the edge of the wave-front and may lead to measurable effects. Every point at the edge of a wave-front generates continuously out-going fully spherical waves propagating at the same phase velocity. The instant position of all wave-fronts at an increment of the wavelength is schematically shown in Fig. 1. Due to the exact rotational symmetry of the system, diffraction at the edge represents movement equally to all directions and, to some extent, the particle therefore undergoes a kind of reflection. As a result, the particle should in principle losses continuously part of its kinetic energy associated with the forward-going wave-front when it travels in a space at a non-zero temperature.

Semi-spherical wavelets generated from the wave-front representing forward-going of the particle remain to be coherent even at finite temperatures and, as the consequence of mutual interference, result in a plane wave which can still be described by quantum mechanics. Let $E_k(x)$ denote the kinetic energy associated with the forward-going wave-front. Diffraction at the edge is supposed to span about one wavelength. Under the condition that $a_0 >> \lambda$, $E_k(x)$ satisfies the equation

$$\frac{dE_k(x)}{E_k(x)} = -\frac{2\pi a_0 \lambda}{\pi a_0^2} \frac{dx}{L} \qquad (2)$$

where $L$ is a temperature dependent coefficient, of dimension length, introduced to ensure the equality. The solution is

$$E_k(x) = E_0 \exp(-2x\lambda/a_0 L) \qquad (3)$$

The lose in $E_k$ in the volume $dx$ due to diffraction at the edge is therefore rewritten as

$$dE_k = -(2\lambda E_0 / a_0 L)\exp(-2x\lambda / a_0 L)dx \qquad (4)$$

when the forward-going plane-wave front is at the position $x$. This is just the energy for the source at the edge to generate spherical waves. The probability for the particle to be at the energy is

$$P_E(x)dx = (2\lambda / a_0 L)\exp(-2x\lambda / a_0 L)dx \qquad (5)$$

Diffraction at the edge of the wave-front generates a series out-going fully spherical waves, i.e., a series of possible states. In principle, the particle may be in any of these states whose energy and probability are well specified. In such a way, statistical physics comes into play and, as a result, transfer of heat becomes possible between the particle system and the heat reservoir. For a real free particle these states form into a continuum and constitute a thermodynamic system in which no mechanical work is involved. The probability of a state at given energy here plays the same role as the degeneracy of an energy level in thermodynamics.

Our assumption is that the thermodynamic part of the particle is in thermal equilibrium with the reservoir all the time when it travels. Please note the following two facts to understand the assumption. The first one is that there is no collision in the whole process and diffraction at the edge involves only a small portion of the plane-wave at a given time. The second one is that a state characterized with a fully spherical wave is a born "thermodynamic state", i.e., a free particle having constant kinetic energy may move equally to all directions. The partition function is then

$$Z(t) = \sum_{n=0}^{Vt/\lambda} \lambda P_E(n\lambda) \exp[-\beta E_0 \lambda P_E(n\lambda)]$$

$$= \int_0^{Vt} P_E(x) \exp[-\beta E_0 \lambda P_E(x)] dx \qquad (6)$$

$$= [e^{-Z_0 \exp(-Z_0 Vt/\lambda)} - e^{-Z_0}]/Z_0$$

Here a constant $Z_0 = \beta h^2/(m a_0 L)$ is introduced, and the notation $\beta = 1/k_B T$ is used as usual with $k_B$ being the Boltzmann constant. The average thermal energy for the thermodynamic part is of the form

$$E_T(t) = U(t)[1 - \exp(-2V\lambda t/a_0 L)] \qquad (7)$$

where the factor $1-\exp(-2V\lambda t/a_0 L)$ is the total probability for the particle in states arising from waves diffracted at the edge, and the function $U(t)$ follows directly from the partition function as in standard thermodynamics[6]

$$\frac{U(t)}{k_B T} = -\beta \frac{\partial \ln Z(t)}{\partial \beta}$$

$$= 1 - \frac{Z_0 + (Z_0^2 N - Z_0) e^{-Z_0[N + \exp(-Z_0 N) - 1]}}{e^{-Z_0[\exp(-Z_0 N) - 1]} - 1} \qquad (8)$$

where $N = Vt/\lambda = E_0 t/h$ is the distance the wave propagated in units of the wavelength. For large $t$, the probability of the part described by quantum mechanics is vanishingly small and the behavior particle of the particle is dominantly described by statistical physics. The energy of the one-dimensional free particle is $k_B T/2$ in the limit of $t \to \infty$, as a result of heat transfer with the reservoir. Consequently, Eq.(8) takes a simpler form and leads to

$$\exp(Z_0) - 2Z_0 - 1 = 0 \qquad (9)$$

The equation may be solved with the help of the Lambert W function[7]. The nonzero real solution can be expressed analytically as

$$Z_0 = -W_{-1}(-\frac{1}{2e^{1/2}}) - \frac{1}{2} \qquad (10)$$

Approximately, we use $Z_0=1.25643$ in numerical calculations.

It is appealing that a quantum particle at a finite temperature can be scaled in the way as shown above so that the whole time dependence is obtained and all properties are expressed only in terms of the temperature and elementary quantities of the single particle. Hereafter, parameters $a_0$ and $L$ do not appear any longer in formalism. For example, Eq. (3) is rewritten as

$$E_k(x) = E_0 \exp(-\frac{Z_0 k_B T}{E_0 \lambda} x) \qquad (11)$$

The probability function defined in Eq.(5) is rewritten as

$$P_E(x) = \frac{Z_0 k_B T}{E_0 \lambda} \exp(-\frac{Z_0 k_B T}{E_0 \lambda} x) \qquad (12)$$

Furthermore the function $U(t)$ becomes a universal function for a continuum of diffracted states. In Fig.2, its behavior is plotted as a function of time elapsed. After an initial oscillation, the function shows little time dependence for large $t$.

At finite temperatures, since the particle can transfer heat with its environment, spherical waves due to diffraction at the edge are no longer coherent. Because an out-going spherical wave is uniformly distributed on its sphere, the overall probability projected on *x*-axis can be calculated without mathematical difficulties. In the range $-Vt<x<Vt$, the probability density of the particle is

$$P(x,t) = \frac{Z_0 k_B T}{2E_0 \lambda} \{Ei(-\frac{Z_0 k_B T t}{h}) \\ - Ei[\frac{Z_0 k_B T}{2E_0}(\frac{x}{\lambda} - \frac{E_0 t}{h})]\} \qquad (13)$$

where $Ei(z)$ is the exponential integral function with argument $z$.[8] It represents a kind of wave packet, sharply peaked near $x=Vt$, exhibiting a continuous spreading. It is true that part of the spreading wave propagates from right to the left, as if the particle is partly reflected. The integration

$$\int_{-Vt}^{Vt} P(x,t)dx = 1 - \exp(-Z_0 k_B T t/h) \qquad (14)$$

determines the total probability at time $t$ for the particle in the thermodynamic part in accordance with Eq. (7).

Since the system of the particle is in thermal equilibrium with the heat reservoir and no mechanical work is involved in the process, the thermodynamic identity is then[6]

$$k_B T dS = [1 - \exp(-Z_0 k_B T t/h)]dU \qquad (15)$$

from which the entropy of the particle can be found

$$S(t) = [1 - \exp(-Z_0 k_B T t/h)]\frac{U(t)}{k_B T} \qquad (16)$$

Considering the behavior of the function $U(t)$, the entropy from each degree freedom of the particle is simply $k_B/2$ in the limit $t\to\infty$. Therefore, it can be conclude that if independent motion is allowed for each freedom, the entropy of a three-dimensional free quantum particle at constant temperature will eventually converge at $3k_B/2$.

We have known probability for the part described by quantum mechanics and that by statistical mechanics. The expectation value of the total energy for the particle is

$$E(t) = E_0 \exp(-Z_0 k_B T t / h) + U(t)[1 - \exp(-Z_0 k_B T t / h)] \qquad (17)$$

The first term is the quantum mechanical part comes from the plane-wave form forward-going wave-front of the particle, and the second term is the thermodynamic part comes from all those waves diffracted at the edge of the wave-front. A particle freely moving in a space at a finite temperature, whose kinetic energy is larger then the average thermal energy, undergoes a continuous loss of its kinetic energy and gives out heat. The conclusion is a consequence of the wavelike nature of the particle and is purely a quantum mechanical feature without classical counterpart. Another outcome of theoretical interest is that a quantum particle originally at rest, if there exist any kind of disturbances or fluctuations, will absorb heat from the environment and eventually possess a thermal energy as described in thermodynamics.

In conclusion, although effects arising form the potential of confinement and interaction between particles in a real system are not considered, the simple model here already shows the possibility that a single quantum particle feels the temperature of the space without need to ascribe a temperature to it. The underlying physics is diffraction at the edge of the wave front of its matter wave.

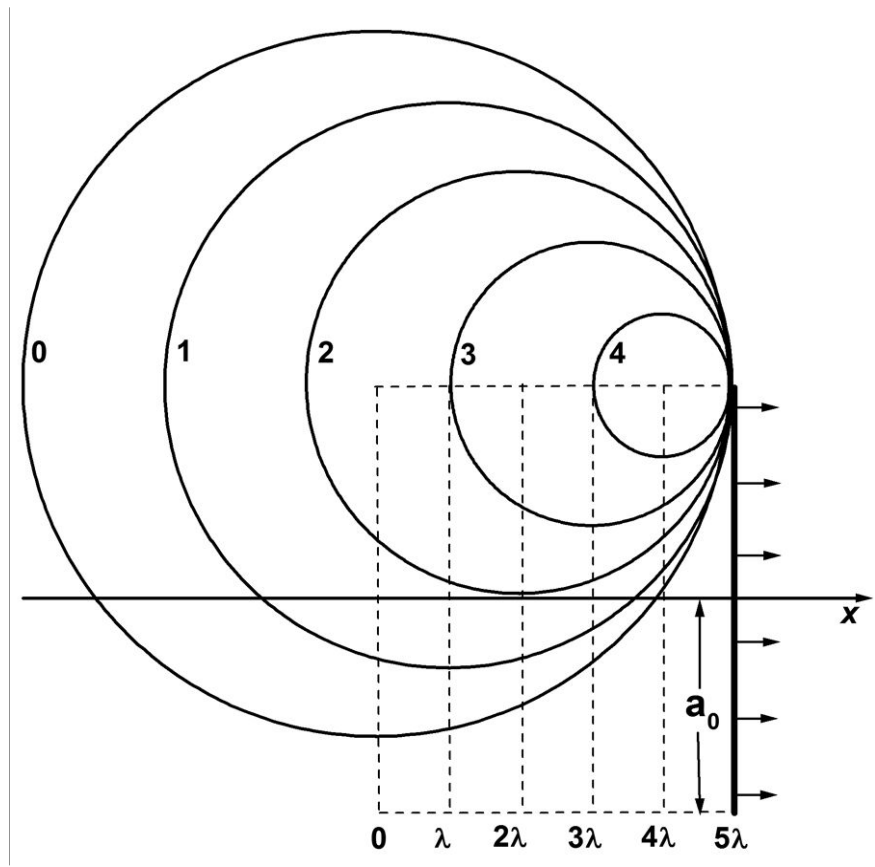

Fig. 1. Schematic for the instantaneous position of the wave fronts of the forward-going plane-wave and spherical out-going waves due to diffraction at the edge. We show here only those waves generated from the up-most point of the edge at an increment of the wavelength in space. The spherical wave labeled $n$ originated from diffraction when the forward-going wave front was at $x=n\lambda$.

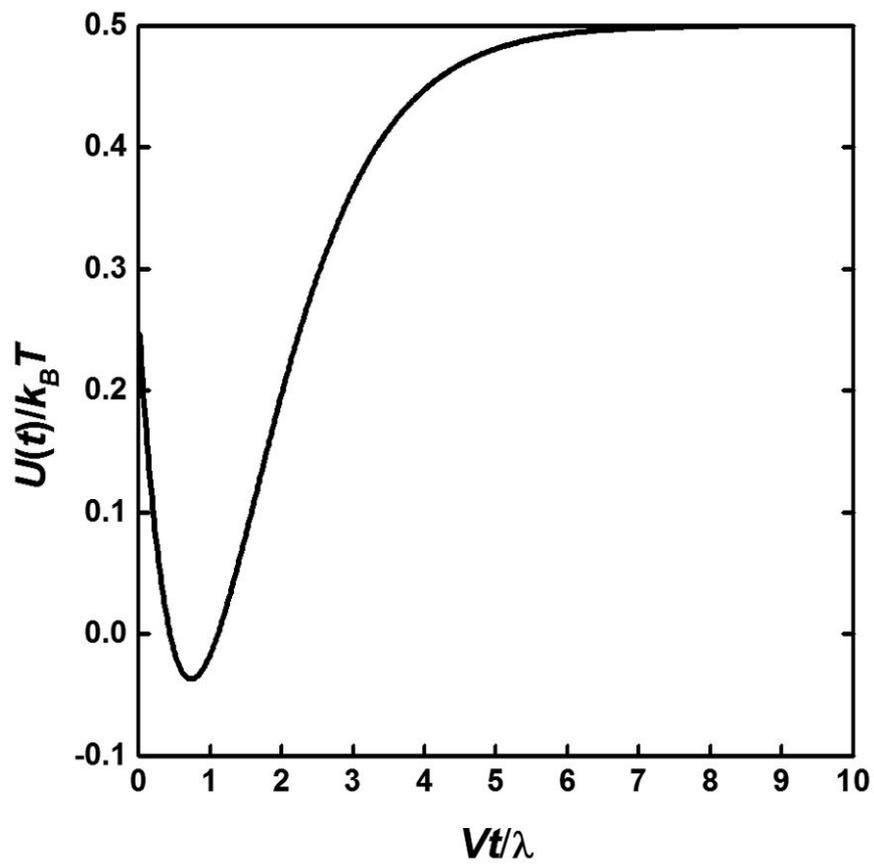

Fig. 2. Behavior of $U(t)$ as a function of time.